\documentclass[sigconf,9pt]{acmart}
\copyrightyear{2026}
\acmYear{2026}
\setcopyright{cc}
\setcctype{by}
\acmConference[ICCAD '26]{IEEE/ACM International Conference on Computer-Aided Design}{November 08--12, 2026}{San Jose, CA, USA}
\acmBooktitle{IEEE/ACM International Conference on Computer-Aided Design (ICCAD '26), November 08--12, 2026, San Jose, CA, USA}
\acmDOI{10.1145/3831252.3833976}
\acmISBN{979-8-4007-2873-0/2026/11}

\usepackage{hhline}
\usepackage{colortbl}
\usepackage{centernot}
\usepackage{pbox}
\usepackage{amsmath}
\usepackage{amsfonts}
\usepackage{url}
\usepackage{bm}
\usepackage{comment}
\usepackage{graphicx}
\usepackage{subfigure}
\usepackage{epstopdf}
\usepackage{multirow}
\usepackage{color}
\usepackage{tabu}
\usepackage{mdframed}
\usepackage{syntax}
\usepackage{fancyvrb}
\usepackage{cleveref}

\usepackage[ruled,linesnumbered,algo2e]{algorithm2e}
\usepackage[noend]{algpseudocode}
\usepackage{algorithmicx}
\usepackage{algpseudocode}
\usepackage{paralist}
\usepackage{stmaryrd}
\usepackage{tikz}
\usepackage[referable]{threeparttablex}
\usepackage{tabularx}
\usepackage{booktabs}
\usepackage{array}
\usepackage{fancyhdr}
\usepackage{float}
\usepackage{xspace}
\usepackage{pifont}
\usepackage{balance}
\usepackage{xcolor}% http://ctan.org/pkg/xcolor

\usepackage{tablefootnote}
\usepackage[normalem]{ulem}
\usepackage{bbding}

\usepackage{tikz}
\usepackage{xcolor}

\usepackage{graphicx}
\usepackage{adjustbox}

\usepackage{float}

\newfloat{figtab}{htb}{fgtb}
\makeatletter
  \newcommand\figcaption{\def\@captype{figure}\caption}
  \newcommand\tabcaption{\def\@captype{table}\caption}
\makeatother

\definecolor{princetonorange}{RGB}{255,143,0}

\definecolor{lightgreen}{RGB}{198, 224, 183}
\definecolor{lightred}{RGB}{240, 205, 176}
\definecolor{USTgold}{RGB}{153,102,0}
\definecolor{USTyellow}{RGB}{204,153,0}
\definecolor{USTyellowlight}{RGB}{255,212,0}
\definecolor{USTorange}{RGB}{255,166,26}
\definecolor{USTpink}{RGB}{255,157,157}
\definecolor{USTblue}{RGB}{0,51,102}
\definecolor{USTmiddle}{RGB}{0,116,188}
\definecolor{USTlight}{RGB}{99,202,225}
\definecolor{USTgray}{RGB}{204,204,204}
\definecolor{USTred}{RGB}{237,27,47}
\definecolor{USTdarkred}{RGB}{124,35,72}

\usepackage{tcolorbox}
\tcbuselibrary{skins,breakable}
    {\endtcolorbox}

\usepackage{geometry}
\begin{document}

% \title{Slack Directly Annotated on Your Verilog: \\Fine-Grained RTL Timing Evaluation for Early Optimization}

% \vspace{-.2in}

\title{E-Layout: Layout-Stage Logic Restructuring via RL-Optimized\\ Equality-Saturated Graph Retrieval-Augmented Generation}

\author{Wenji Fang$^{1\dagger}$,
Rongjian Liang$^{2}$,
Yi-Chen Lu$^{3\dagger}$,
Cunxi Yu$^{2,4}$, \\
Chen Chen$^{4}$, 
Shang Liu$^{1}$,
Chenhui Deng$^{2}$,
Wenkai Li$^{1}$,
Haoxing Ren$^{5\dagger}$,
Zhiyao Xie$^{1*}$}
% \authornote{$^*$ Corresponding author (eezhiyao@ust.hk).}
% \footnote{ Corresponding Author}
\affiliation{%
  \institution{$^{1}$Hong Kong University of Science and Technology}
  \institution{$^{2}$NVIDIA $^{3}$Ricursive Intelligence $^{4}$University of Maryland, College Park $^{5}$Agentrys}
  % \institution{}
  % \institution{}
  \country{}
}
% \renewcommand{\thefootnote}{\fnsymbol{footnote}}
% \footnotetext[2]{Work done while Wenji Fang was an intern at NVIDIA.}
% \footnotetext[1]{Corresponding author (eezhiyao@ust.hk).}
% \renewcommand{\thefootnote}{\arabic{footnote}}

% \renewcommand{\thefootnote}{\fnsymbol{footnote}}
% \footnotetext[2]{Work done while Wenji Fang was an intern at NVIDIA.}
% \footnotetext[1]{Corresponding Author (eezhiyao@ust.hk)}
% \renewcommand{\thefootnote}{\arabic{footnote}}

\pagestyle{empty}

\begin{abstract}

Achieving high sign-off performance in advanced ICs increasingly depends on layout-stage optimization under realistic physical conditions. While recent efforts have moved toward synthesis–layout fusion, most existing works still focus on bringing estimated physical information earlier into synthesis, whose optimization does not fully correlate with real layout-stage behavior. Meanwhile, layout-stage optimization remains largely restricted to a fixed logic structure, relying mainly on buffering and sizing.
This fundamentally constrains the optimization space under realistic physical conditions.
To address this, we present E-Layout, an ML-driven framework for post-placement layout restructuring.
E-Layout jointly optimizes logic transformation and physical implementation across multiple subcircuits under design-level timing feedback. This enables the selection of the most timing-beneficial set of disjoint restructured subcircuits to form the optimized circuit.
Specifically, it first constructs equivalence-preserving logic alternatives via e-graph equality saturation, then couples logic selection with physical realization through a graph-based retrieval-augmented generation (RAG) model, and finally refines these local logic–physical decisions with reinforcement learning under design-level timing feedback.
Evaluated on 9 diverse designs, it achieves on average 6\% TNS improvement and 2\% WNS improvement, while even reducing area by 1\%, over an industry-leading commercial physical design flow.\looseness=-1

\end{abstract}

\maketitle

\begingroup
\renewcommand{\thefootnote}{\fnsymbol{footnote}}
\footnotetext[2]{The work was conducted while Wenji Fang was an intern at NVIDIA, and Yi-Chen Lu and Haoxing Ren were at NVIDIA.}
\footnotetext[1]{Corresponding author (eezhiyao@ust.hk).}
\endgroup

% \vspace{-.1in}
\section{Introduction}\label{sec:intro}

Modern VLSI design flows have traditionally treated logic synthesis and physical design as largely separate stages. Logic synthesis determines most of the gate-level structure early, while downstream layout stages primarily refine the physical implementation through placement, routing, and related optimizations, such as buffering and gate sizing. Although this separation simplifies flow organization and EDA tool development, it also fixes many logic structural decisions before accurate physical information becomes available. As a result, optimization opportunities that depend on real placement and interconnect context are often difficult to fully exploit in later physical stages.

\textbf{The trend of synthesis--layout fusion.}
In recent years, this rigid boundary has gradually blurred. In industry, commercial EDA tools increasingly emphasize tighter cross-stage integration to improve optimization at advanced technology nodes~\cite{synopsys,cadence}. In academia, this trend has been explored mainly through physical-aware synthesis~\cite{gao1993placement,liu2011simultaneous,liu2023aimap,pan2024physically}, where estimated placement and parasitic information are fed back into earlier synthesis stages to guide iterative resynthesis. Together, these efforts reflect a broader shift from stage-isolated optimization toward logic--physical co-optimization.\looseness=-1

\begin{figure}[!t]
  \centering
  % \vspace{-.1in}
  \includegraphics[width=1\linewidth]{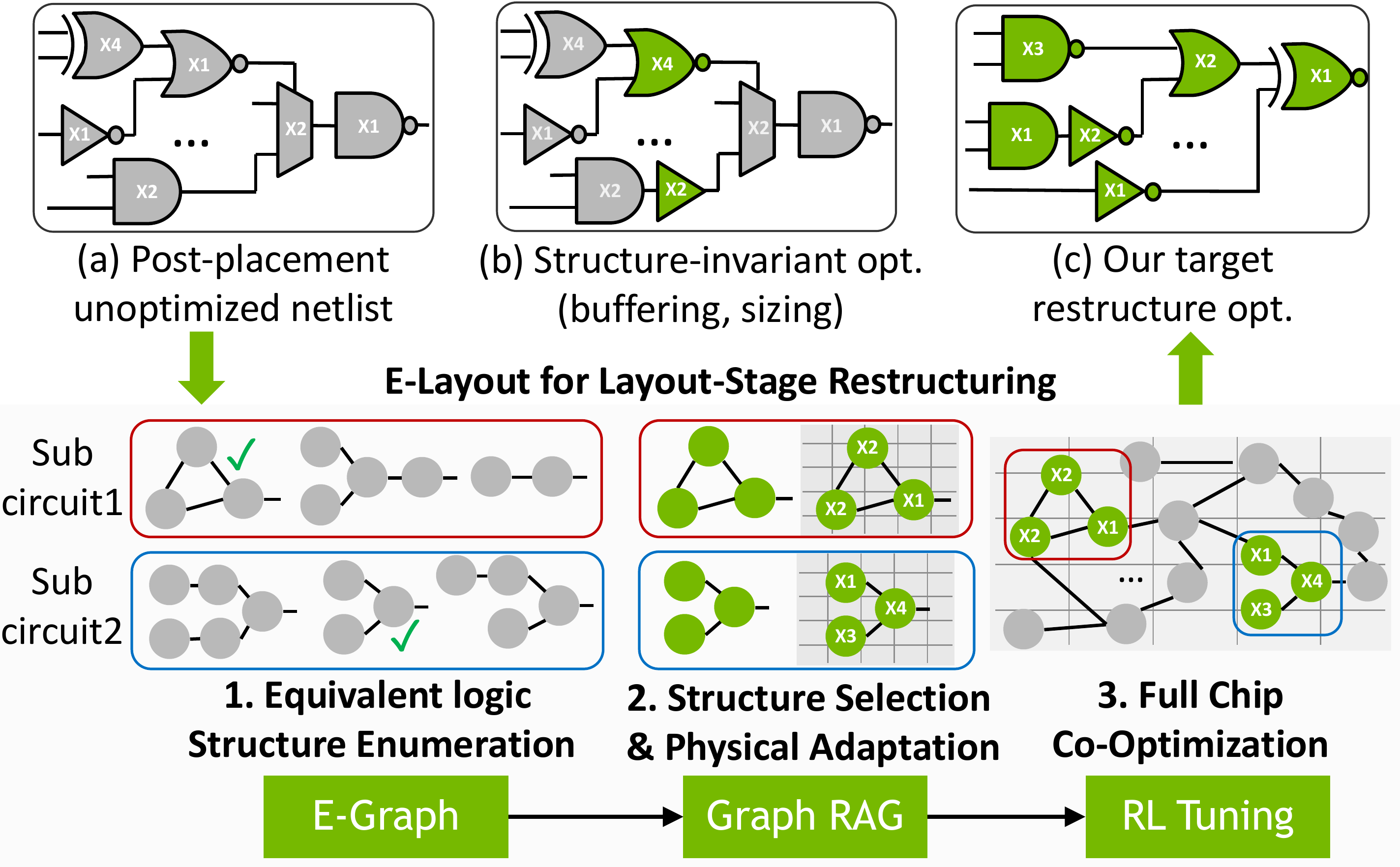}
  \vspace{-.1in}
  \caption{Overview of E-Layout. E-Layout targets layout-stage restructuring, opening a much larger optimization space than structure-invariant methods. It decomposes the task into three coupled steps: equivalent structure exploration, logic selection with physical realization, and design-level timing-driven refinement across disjoint subcircuits.}
  \label{fig:motivation}
  \vspace{-.3in}
\end{figure}

\textbf{The current academic focus: physical-aware synthesis.}
Despite this broader trend, most recent academic efforts have focused on bringing physical information earlier into synthesis, rather than enabling direct restructuring later at the layout stage. This distinction is important: physical-aware synthesis operates before final placement is available and therefore relies on approximate physical estimates. While such estimates can guide early logic optimization, they do not fully capture the true post-placement timing behavior of the final implementation, and may therefore lead to restructuring decisions that deviate from the actual optimum.

\textbf{Layout-stage optimization.}
This limitation naturally shifts the focus toward the layout stage, where direct restructuring can be performed under the real physical context, as shown in~\Cref{fig:motivation}. At this stage, gate locations and interconnect context are available, enabling more accurate timing correlation.
However, optimization at this stage remains dominated by \emph{structure-invariant} techniques, such as buffering~\cite{chu2007flute,alpert2006accurate,van1990buffer,wu2024aito,liang2023bufformer,hsiao2025buffalo,kahng2025recursive,lu2021rl} and gate sizing~\cite{ning1994strongly,zhou2006gate,cheng2023dagsizer,nath2022transsizer,lu2025lego,ye2024learning}, which refine the physical implementation without changing the underlying logic structure. This fundamentally limits the optimization space at the layout stage.

\textbf{Limitations of existing layout restructuring.}
To enlarge the optimization space beyond structure-invariant buffering and sizing, a limited number of earlier works~\cite{pedram1991layout,jiang1997post, chang1997postlayout, kim2008framework} also explored layout-stage restructuring decades ago. These studies investigated layout-driven restructuring through placement-aware decomposition, incremental rewiring, and layout-level remapping. However, they remained largely heuristic and locally scoped: the restructuring space was limited to narrow rewiring or remapping opportunities on local subcircuits, and physical realization was not jointly optimized with logic transformation.

This limitation also persists in modern, widely adopted open-source tools. For example, OpenROAD~\cite{ajayi2019openroad} identifies timing-critical regions and invokes ABC~\cite{brayton2010abc} for logic remapping. However, because ABC operates purely at the logic level without physical awareness, the resulting remapping can significantly break physical coherence and lead to weak post-placement correlation.
OpenPhySyn~\cite{agiza2020openphysyn} performs physically aware local edits such as pin swapping and gate cloning, together with timing-repair operations such as buffer insertion and resizing, but these cell-level transforms offer only limited restructuring capability and cannot explore meaningfully different logic structures. Consequently, modern open-source layout-stage restructuring remains either loosely coupled to physical context or restricted to narrow local edits.

\begin{table}[!t]
% \vspace{-.35in}
\caption{Comparison with existing layout-stage logic restructuring methods.\looseness=-1}%applied in optimization.}
\vspace{-.15in}
\resizebox{0.5\textwidth}{!}{ 

\begin{tabular}{cc||c|c|c} \toprule
\multicolumn{2}{c||}{\textbf{Feature}}                                                                    & \textbf{\begin{tabular}[c]{@{}c@{}}OpenROAD \\      Restructure~\cite{ajayi2019openroad}\end{tabular}} & \textbf{\begin{tabular}[c]{@{}c@{}}OpenPhySyn \\      Restructure~\cite{agiza2020openphysyn}\end{tabular}} & \cellcolor[HTML]{cde09a}\textbf{\begin{tabular}[c]{@{}c@{}}E-Layout \\      (ours)\end{tabular}} \\ \midrule
\multirow{3}{*}{\textbf{Rewrite}}                                                  & Method          & re-mapping                                                                    & \begin{tabular}[c]{@{}c@{}}pin-swap, \\      gate clone\end{tabular}      & \begin{tabular}[c]{@{}c@{}}e-graph +\\      graph RAG\end{tabular}       \\ \cline{3-5}
                                                                                       & Scope                      & subcircuit                                                                           & cell                                                                            & subcircuit                                                                      \\ 
                                                                                       & Logic change                & \Checkmark                                                                           & \XSolidBrush                                                                              & \Checkmark                                                                      \\ \midrule
\multirow{3}{*}{\textbf{\begin{tabular}[c]{@{}c@{}}Opt   \\      Impact\end{tabular}}} & Design timing              & \XSolidBrush                                                                            & \Checkmark                                                                             & \Checkmark                                                                      \\
                                                                                       & Delay target         & ABC (unit delay)                                                                           & layout (NLDM)                                                                          & layout (NLDM)                                                               \\
                                                                                       & Opt. space                 & large                                                                         & small                                                                           & medium                                                     \\ \bottomrule              
\end{tabular}

}
\label{tbl:baseline}
\vspace{-.05in}
\end{table}

\begin{figure}[!t]
  \centering
  % \vspace{-.35in}
  \includegraphics[width=1\linewidth]{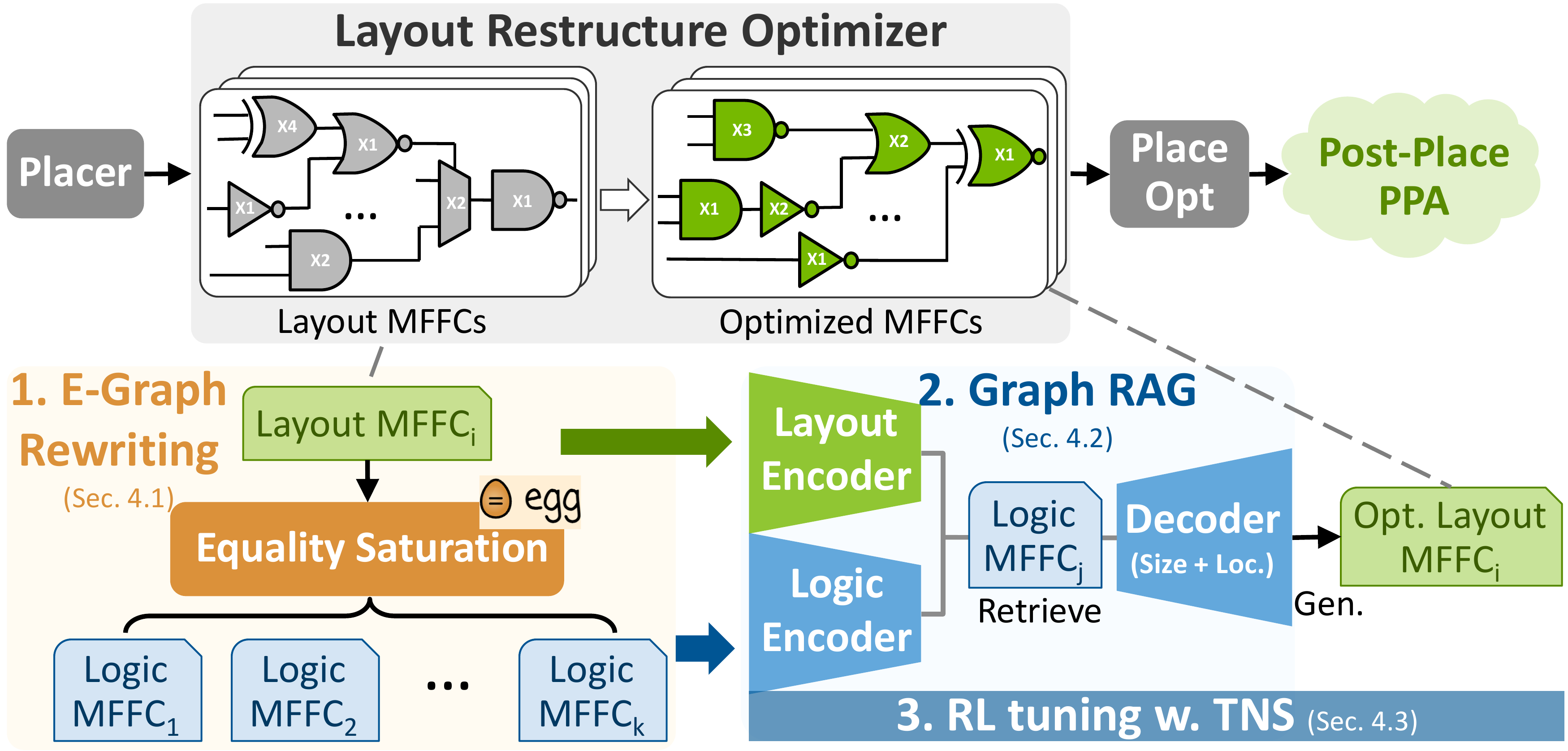}
  \vspace{-.2in}
  \caption{Workflow of E-Layout and its integration into the placement flow. E-Layout first uses e-graph equality saturation to provide functionally equivalent logic MFFC candidates, then applies an RL-optimized graph RAG model to select new structures and generate gate sizes and locations.\looseness=-1} 
  \label{fig:overview}
  \vspace{-.1in}
\end{figure}

\textbf{Our solution: post-placement logic--layout co-optimization.}
To address these limitations, we propose \textbf{E-Layout}, an ML-driven framework for post-placement\footnote{We choose post-placement because it is the earliest stage with real physical information. Earlier synthesis stages rely on estimates, whereas later sign-off and ECO stages typically offer a smaller optimization space. In principle, the framework can also be extended to other stages.} layout restructuring, as shown in~\Cref{fig:overview}. The key idea is to jointly optimize logic structure transformation and physical adaptation (i.e., resize and relocate for each new cell) across multiple subcircuits under global design-level timing feedback. As a result, the most timing-beneficial set of disjoint restructurings can be selected to form the optimized circuit.\looseness=-1

Specifically, E-Layout operates at the subcircuit level to balance rich restructuring opportunities with physically coherent local updates, while jointly optimizing multiple local subcircuits under global design timing feedback.
It decomposes layout restructuring into three steps, each enabled by a dedicated strategy:\looseness=-1
\begin{enumerate}
    \item \textbf{Equivalent structure exploration.} Layout-stage restructuring should rewrite the post-synthesis logic structure while strictly preserving functional equivalence. To achieve this, E-Layout uses e-graph equality saturation to expand each local subcircuit into a compact space of functionally equivalent pure logic structures. This enables much broader structural exploration than isolated local rewrites while preserving correctness by construction.
    \item \textbf{Logic selection with physical realization.} A promising logic structure is useful only after it is physically realized and evaluated under the surrounding layout context. To this end, we develop a graph retrieval-augmented generation (RAG) model to retrieve promising structures from the e-graph rewritten candidate space and jointly generate physically consistent gate sizes and locations, thereby coupling logic selection with physical adaptation.
    \item \textbf{Design-level timing refinement.} Promising local subcircuit restructurings should not be judged in isolation, because their timing impact depends on full-chip interaction across all subcircuits in the design. To this end, E-Layout further tunes the graph RAG model with reinforcement learning (RL) using full-chip timing feedback, refining local logic--physical decisions toward improved design-level post-placement timing.
\end{enumerate}

Evaluated on 9 diverse designs, E-Layout delivers on average 6\% TNS improvement and 2\% WNS improvement, while even achieving 1\% area reduction over a commercial physical design flow that has already undergone extensive optimization. We further demonstrate its impact on critical paths at both the design and subcircuit levels. E-Layout is also designed for scalability: e-graph rewriting is applied to non-overlapping subcircuits that can be processed in parallel, while graph RAG inference in each RL iteration completes within seconds. Together, these results highlight the promise of combining symbolic reasoning with ML techniques for layout-stage restructuring optimization.

% \section{Preliminaries and Problem Formulation}\label{sec:pre}

\begin{figure*}[!t]
  \centering
  % \vspace{-.4in}
  \includegraphics[width=1\linewidth]{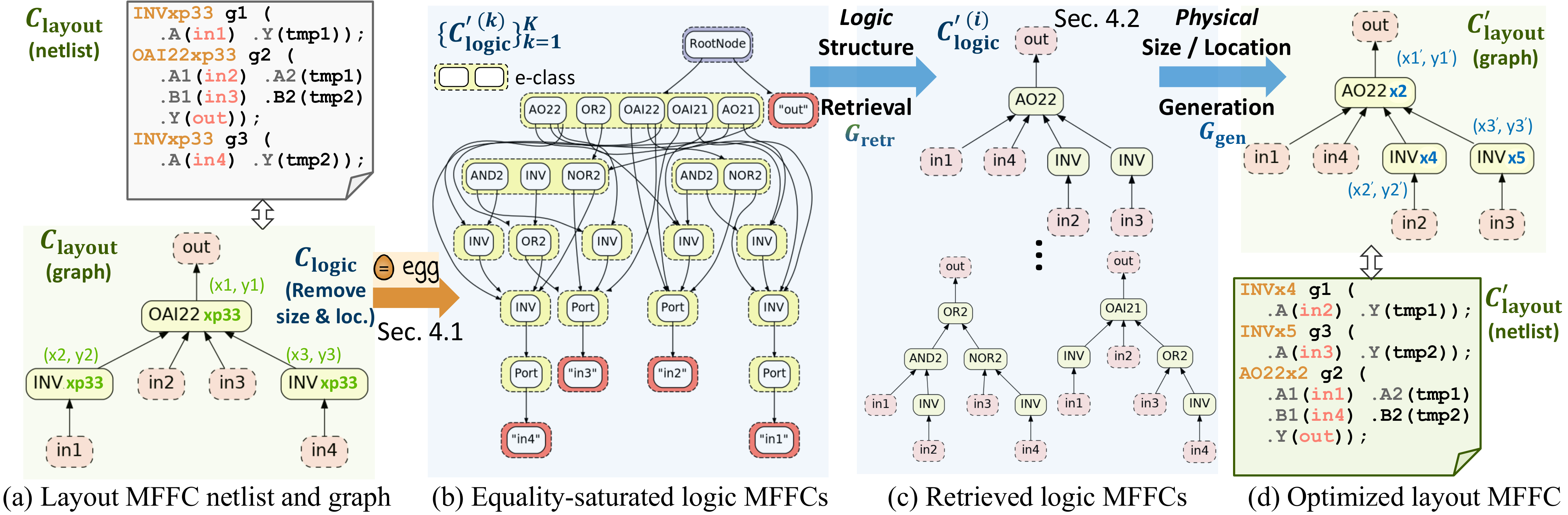}
  \vspace{-.25in}
  \caption{E-Layout restructuring flow in the MFFC perspective. (a) Extract a post-placement layout MFFC graph. (b) Apply e-graph equality saturation to enumerate equivalent logic structures. (c) Retrieve logic structure candidates, then (d) Generate physical sizes and locations to form an optimized layout MFFC to substitute the original one.\looseness=-1} 
  \label{fig:egraph}
  \vspace{-.05in}
\end{figure*}

\section{Preliminaries}\label{sec:prelim}

% \Cref{fig:overview} illustrates how our E-Layout integrates into a standard commercial physical design flow. 

\subsection{MFFC in Circuit Optimization}
A \emph{maximum fanout-free cone} (MFFC) is a common structural abstraction in logic synthesis, including both logic rewriting and technology mapping~\cite{brayton2006scalable,chen2025revisit}. Formally, the MFFC of a node is the largest subset of nodes such that every path from any node in the subset to the primary outputs passes through that node. Intuitively, it contains all logic used exclusively to drive that node, and can therefore be replaced as a whole when the node is substituted. This makes MFFCs a natural unit for local circuit optimization, since transformations within an MFFC do not directly affect other parts of the circuit through shared internal logic. In E-Layout, we likewise adopt MFFCs as the basic subcircuit granularity for layout restructuring.

\subsection{E-Graphs for Circuit Optimization}
An e-graph is a compact data structure for representing many equivalent expressions simultaneously, and has been increasingly adopted in circuit optimization.
E-graphs are typically used in a two-stage workflow: \emph{equality saturation} followed by \emph{extraction}. During equality saturation, rewriting rules are repeatedly applied to expand the e-graph with functionally equivalent circuit structures. Once saturation reaches a predefined resource limit or fixed point, an optimized implementation is extracted according to a target cost function, such as area or delay. By avoiding early commitment to a single rewriting sequence, e-graphs enable efficient exploration of large equivalence spaces and have been applied to a variety of circuit representations, including RTL~\cite{coward2024rover,coward2023automating,zhang2025aspen,liu2025efficient}, AIGs~\cite{chen2024syn,chen2025morphic}, and HLS~\cite{cheng2024seer,ustun2023equality}. Despite their success in logic optimization, standard e-graphs are designed for pure logic representations and therefore do not directly support layout-stage restructuring with physical attributes and post-placement timing.

% \subsection{RAG and RL in LLM}
% Retrieval-augmented generation (RAG) decomposes generation into two stages: retrieving promising candidates from an external pool, and generating outputs conditioned on the retrieved candidates. This paradigm is attractive when the output space is large but can be constrained by a structured candidate set. Reinforcement learning (RL) is often used to further optimize generation policies when the target objective is defined by end-task feedback rather than direct supervision. By treating generation as a policy and downstream quality as reward, RL enables candidate selection and generation decisions to be refined toward task-level objectives.

\section{Problem Formulation}\label{sec:prob}

Given a placed but unoptimized layout $\mathcal{G}$, our goal is to perform post-placement restructuring and obtain a restructured layout $\mathcal{G}'$ that improves timing while preserving functional equivalence. After restructuring, $\mathcal{G}'$ is passed back to the commercial physical design tool for incremental legalization and subsequent structure-invariant optimizations such as buffering and sizing. The final quality of the restructured design is then evaluated using timing and area metrics, including total negative slack (TNS), worst negative slack (WNS), and area. Formally, we seek
\[
\mathcal{G}' = \arg\min_{\hat{\mathcal{G}} \in \Omega(\mathcal{G})}
\; \mathcal{L}_{\text{PPA}}(\hat{\mathcal{G}})
\quad
\text{s.t.}
\quad
\hat{\mathcal{G}} \equiv \mathcal{G},
\]
where $\Omega(\mathcal{G})$ denotes the set of physically realizable restructured layouts derived from $\mathcal{G}$ through layout-stage restructuring, $\hat{\mathcal{G}} \equiv \mathcal{G}$ enforces functional equivalence, and $\mathcal{L}_{\text{PPA}}$ is an optimization objective defined over post-layout metrics, including TNS, WNS, and area.

Specifically, layout-stage restructuring should jointly optimize two coupled aspects:
(1) \emph{logic transformation}, by replacing local subcircuits with functionally equivalent alternatives; and
(2) \emph{physical adaptation}, by assigning physically consistent attributes to the transformed structure, including gate sizes and placement locations of each cell.

\section{Methodology}\label{sec:method}

% \subsection{Method Overview.} 

% In~\Cref{sec:egraph}, we will first introduce the e-graph rewriting for layout standard cell graph, and then in~\Cref{sec:rag}, we will elaborate the model architecture of our propsed graph RAG model, and finally in~\Cref{sec:rl}, we will introduce the RL tuning method.

The overall framework of E-Layout is shown in~\Cref{fig:overview}. E-Layout is built on the view that layout-stage restructuring should couple \emph{logic transformation} and \emph{physical implementation}, and optimize them under \emph{design-level timing feedback}. Accordingly, it proceeds in three tightly connected stages. First, as described in~\Cref{sec:egraph}, E-Layout applies e-graph rewriting to layout standard-cell graphs to construct an equivalence-preserving space of candidate logic structures. Second, as illustrated in~\Cref{sec:rag}, it uses a graph RAG model to select promising structures from these alternatives and generate their physical sizes and locations, thereby bridging logic and physical implementation. Third, as detailed in~\Cref{sec:rl}, it further refines these local logic–physical decisions by tuning the graph RAG model with reinforcement learning under layout timing feedback, so that restructuring is ultimately guided by full-chip timing improvement.

% As shown in~\Cref{fig:overview}, within E-Layout, our proposed restructuring framework consists of two steps: (1) E-graph–based layout restructuring for equivalence exploration., and then in (2) Graph RAG with RL for timing optimization.  Using the candidate pool, we develop a graph-based retrieval-augmented generation model that retrieves promising subcircuit logic structures and generates per-cell physical attributes (i.e., sizing and location). The policy is optimized via reinforcement learning with real-time PPA feedback, enabling joint improvement of structure retrieval, sizing, and local placement under the layout timing constraints.The detailed methodology for each of these two stages is described below.

\subsection{Layout MFFC Rewriting via E-Graph Equality Saturation}\label{sec:egraph}

At the first stage of E-Layout, we use e-graph rewriting to expand the structural search space by systematically enumerating functionally equivalent alternatives on layout standard-cell graphs. In this way, E-Layout exposes richer restructuring opportunities while preserving correctness.

However, directly applying the standard e-graph saturation and extraction paradigm to our target layout is not applicable, mainly due to two challenges: 
\begin{itemize}
    \item \textbf{Saturation: difficult on complex standard cells with physical attributes.} Layout circuits contain diverse standard cells with complex logic functions and physical attributes. Existing e-graph rewriting rule sets are mainly designed for pure and simple logic forms (e.g., AIG and RTL), making them hard to reuse for standard-cell graphs.
    \item \textbf{Extraction: lack of PPA-aware cost.}  Conventional e-graph extraction is driven by simple linear cost functions (e.g., node count, logic depth), which cannot capture layout PPA with non-linear, topology- and placement-dependent delays, thus limiting physical-aware logic optimization.
\end{itemize}

% In addressing these challenges, we leverage the e-graph only to generate diverse logically equivalent pure-logic structures through saturation. For the selection of the new structures and decision of new gate physical attributes, our subsequent graph-based RAG model.
As shown in~\Cref{fig:egraph}, to overcome these challenges, we leverage the e-graph only to provide diverse equivalent pure-logic structures via saturation. Then, the extraction of physically optimized layouts, which selects suitable structures and assigns new gate physical attributes, is delegated to our subsequent graph-based RAG model. We detail the e-graph equality saturation process designed for standard-cell layout below.

% However, conventional e-graph frameworks (e.g., egg~\cite{willsey2021egg}) are designed for logic-only graphs and are unaware of physical design constraints such as placement, wirelength, or timing correlation. In E-Layout, 

\textbf{Handling layout at MFFC level.}
As shown in~\Cref{fig:egraph} (a), we perform e-graph saturation at the MFFC granularity, denoted as $C_\text{layout}$. Since no logic inside an MFFC is shared with nodes outside the cone, optimizing an MFFC can directly improve PPA without disturbing the rest of the circuit~\cite{chen2025revisit}. This allows us to explore new structures for multiple nodes within an MFFC independently. Specifically, we extract the MFFC for each gate from the placed, unoptimized layout and represent it as a graph. To enable pure-logic rewriting, we remove all physical attributes (i.e., size, location) from $C_\text{layout}$ and retain only the logical structure for e-graph saturation, denoted as $C_\text{logic}$.\looseness=-1

\begin{table}[!t]

\caption{Standard cell rewriting rules.}%applied in optimization.}
\vspace{-.05in}
\resizebox{0.5\textwidth}{!}{
\begin{tabular}{c||c|c}
\toprule
\textbf{Type} & \multicolumn{2}{c}{\textbf{Standard Cell Rewriting Rules}}                \\ \midrule
\multirow{4}{*}{\begin{tabular}[c]{@{}c@{}} \textbf{Type I: one-level} \\      \textbf{de/recompose}\\      \textbf{for complex gates}\end{tabular}} & \multicolumn{2}{c}{$\mathrm{AO22} \Leftrightarrow\ \mathrm{AO21} (x_1, x_2,   \mathrm{AND2} (y_1, y_2))$}                \\
                                                                                                                                  & \multicolumn{2}{c}{$\mathrm{AND3} \Leftrightarrow\ \mathrm{AND2} (\mathrm{AND2}   (x, y), z)$}                          \\
                                                                                                                                  & \multicolumn{2}{c}{$\mathrm{OA22} \Leftrightarrow\ \mathrm{INV} (\mathrm{AO21}   ...)$}                                \\
                                                                                                                                  & \multicolumn{2}{c}{…}                                                      \\ \midrule
\multirow{4}{*}{\begin{tabular}[c]{@{}c@{}}\textbf{Type   II: } \\      \textbf{complex} $\Leftrightarrow$ \textbf{primitive}\end{tabular}}                       & \multicolumn{2}{c}{$\mathrm{AO22} \Leftrightarrow\ \mathrm{OR2}( \mathrm{AND2}(x1,x2), \mathrm{AND2}(y1,y2) )$}                  \\
                                                                                                                                  & \multicolumn{2}{c}{$\mathrm{NAND2} \Leftrightarrow\ \mathrm{INV}(   \mathrm{AND2}(x,y) )$}                             \\
                                                                                                                                  & \multicolumn{2}{c}{$\mathrm{XOR2} \Leftrightarrow\ \mathrm{OR2}(   \mathrm{AND2}(x,\mathrm{INV}(y)), \mathrm{AND2}(\mathrm{INV}(x),y) )$}         \\
                                                                                                                                  & \multicolumn{2}{c}{…}                                                       \\ \midrule
\multirow{4}{*}{\begin{tabular}[c]{@{}c@{}}\textbf{Type   III:} \\      \textbf{Boolean} \\      \textbf{simplification}\end{tabular}}                         & \multicolumn{1}{l|}{\textit{De Morgan}}        & $\mathrm{INV}(\mathrm{AND2}(x,y)) \Leftrightarrow\ \mathrm{OR2}(\mathrm{INV}(x),\mathrm{INV}(y))$  \\
                                                                                                                                  & \multicolumn{1}{l|}{\textit{Double inversion}} & $\mathrm{INV}(\mathrm{INV}(x)) \Rightarrow x$                     \\
                                                                                                                                  & \multicolumn{1}{l|}{\textit{Absorption}}       & $\mathrm{AND2}(x, \mathrm{OR2}(x,y)) \Rightarrow x$                 \\ \cline{2-3}
                                                                                                                                  & \multicolumn{2}{c}{…}                                                    \\ \bottomrule
\end{tabular}

}

\label{tbl:rule}
\vspace{-.05in}
 
\end{table}

\textbf{E-graph saturation with layout rewriting rules.}
To support e-graph rewriting over standard-cell layouts, we design a compact set of equivalence-preserving rules over library cells, as summarized in~\Cref{tbl:rule}. These rules expand the search space through decomposition, recomposition, and Boolean simplification.
Specifically, we propose three types of standard cell rewriting rules: (1) one-level decomposition/recomposition of complex gates, where complex cells (e.g., AO22) are decomposed into smaller cells (e.g., AO21 + AND2) and recomposed back; (2) complex-to-primitive bidirectional conversions, where complex cells (e.g., AO22) are transformed into multiple primitive gates (e.g., AND2, OR2, INV) and vice versa, enabling finer-grained restructuring; and (3) Boolean simplification, where canonical identities (e.g., De Morgan, double inversion removal, absorption) are applied to provide simplified implementations.\looseness=-1

With these customized rules, we construct an initial e-graph from each layout MFFC $C_\text{layout}$ and run equality saturation until reaching resource limits (e.g., iteration bound, e-node budget, or fixpoint) to enumerate all $K$ equivalent logic MFFCs $\{ C_\text{logic}^{\prime(k)} \}_{k=1}^K$. These equivalent logic MFFCs serve the candidate pool for our subsequent ML framework.
During saturation, the rules systematically enumerate alternative implementations while guaranteeing functional equivalence. As shown in~\Cref{fig:egraph} (b), the resulting saturated e-graph compactly contains a rich design space of candidate pure-logic structures for each MFFC, which will later serve as input to our graph-based RAG model for timing optimization.\looseness=-1

% We apply equality saturation to each layout MFFC to exhaustively enumerate all functionally equivalent subcircuits. As shown in Fig. 2(b), E-Layout expands the logic graph into an equality graph (e-graph), where each e-class represents a set of logically equivalent gate configurations. By applying a series of Boolean and structural transformation rules—such as De Morgan’s law, AOI/OAI decomposition, and inverter propagation—the e-graph compactly encodes a rich family of equivalent logic structures, forming a diverse candidate pool for later learning-based optimization.

\subsection{Graph RAG Model Architecture} \label{sec:rag}

\begin{figure*}[!t]
  \centering
  % \vspace{-.35in}
  \includegraphics[width=1\linewidth]{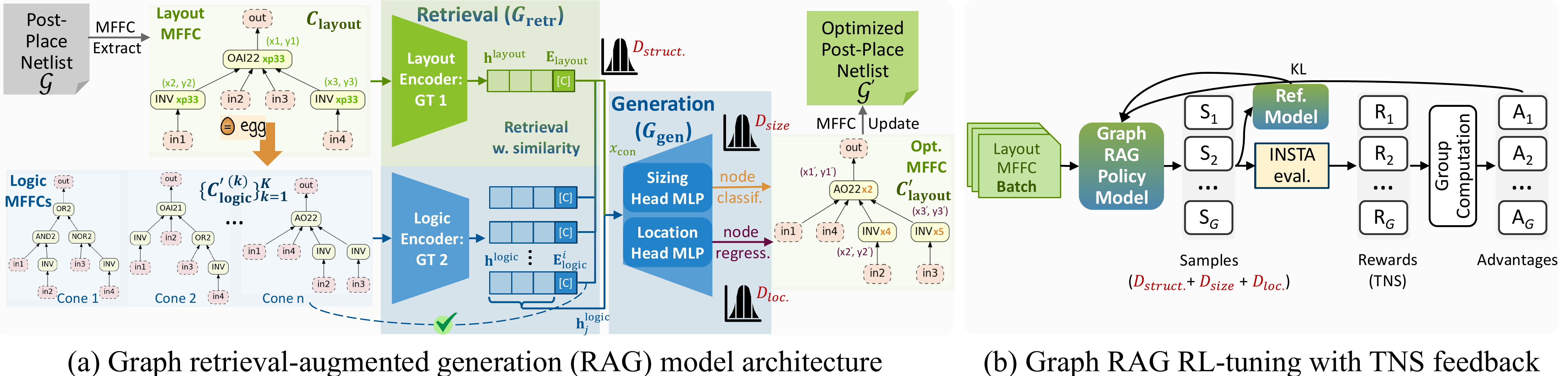}
  \vspace{-.2in}
  \caption{Graph RAG for logic structure retrieval and physical attribute generation, with RL tuning with layout TNS feedback.\looseness=-1}
  \label{fig:rag}
  \vspace{-.05in}
\end{figure*}

% While equality saturation generates a rich set of functionally equivalent MFFCs, it is inherently difficult to directly extract a timing-optimal layout using handcrafted linear cost functions. To overcome this limitation, we replace the traditional extraction step with a graph-based retrieval-augmented generation (RAG) model that learns to select a promising logic structure and generate its corresponding cell-level physical attributes. 

% With these predicted physical attributes, the optimized MFFC layout can be instantiated and evaluated with accurate timing. Then we further propose to tune the graph RAG model using reinforcement learning with the full-chip timing as feedback to improve the timing. In this subsection, we elaborate the graph RAG model architecture, and we illustrate the RL tuning method in~\Cref{sec:rl}

While equality saturation provides a rich space of equivalent MFFCs, selecting a timing-optimal structure cannot be reliably achieved with simple linear extraction costs. To address this, we use a graph-based retrieval-augmented generation (RAG) model to connect structural exploration with physical realization. The equivalent pure-logic candidates $\{ C_\text{logic}^{\prime(k)} \}_{k=1}^K$ serve as the retrieval space, from which the promising structure $C’_{\text{logic}}$ is selected, while gate sizes and locations are generated conditioned on the selected structure and the surrounding layout context. This allows E-Layout to transform equivalence-preserving logic candidates into physically optimized layout subcircuits. As shown in~\Cref{fig:egraph}(c) and (d), the retrieved logic structure $C’_{\text{logic}}$, together with the predicted sizes and locations, forms an optimized $C’_{\text{layout}}$ that replaces the original $C_{\text{layout}}$.

As shown in~\Cref{fig:rag}(a), the graph RAG model consists of two components: (1) a retrieval module $G_{\text{retr}}$, which uses separate graph encoders for the original layout MFFC and the candidate logic MFFCs to retrieve promising structures; and (2) a generation module $G_{\text{gen}}$, which takes the retrieved logic structure together with the original layout context and predicts per-cell sizes and locations. We detail each module below.

\textbf{Logic structure retrieval: dual-tower layout-logic graph encoders.}
The retrieval module  $G_\text{retr}$ contains a layout encoder and a logic encoder. It encodes the original layout MFFC $C_\text{layout}$ and the candidate logic MFFCs $\{ C_\text{logic}^\prime{(k)} \}_{k=1}^K$ into layout embedding $\mathbf{E}_{\text{layout}}$ and logic embedding pool $\{ \mathbf{E}_{\text{logic}}^{(k)} \}_{k=1}^K$, respectively. Then $\mathbf{E}_{\text{layout}}$ serves as the query to retrieve candidates from the logic pool $\{ \mathbf{E}_{\text{logic}}^{(k)} \}_{k=1}^K$ via similarity scoring.

Specifically, the layout encoder maps $C_{\text{layout}}$ into an embedding using both physical and structural information. Given the layout MFFC graph, for each node $n_i$ it combines physical features $\mathbf{x}_i^{\text{phys.}}$ (e.g., cell size, capacitance, slew, slack, and area) and structural features $\mathbf{x}_i^{\text{struct.}}$ (e.g., cell type and number of reachable endpoints) to produce a node embedding $\mathbf{h}_i^{\text{layout}}$. These node embeddings are then aggregated over all nodes to obtain a single cone-level embedding $\mathbf{E}_{\text{layout}}$.
The logic encoder maps $C'_{\text{logic}}$ using structural information only. It operates on equality-saturated candidate MFFCs and uses logical features (e.g., cell type and fanout-of-4 delay) to produce node embeddings $\mathbf{h}_j^{\text{logic}}$, which are similarly aggregated into a graph-level embedding $\mathbf{E}_{\text{logic}}$.
Both encoders are implemented as graph transformers followed by pooling to obtain the graph-level embedding, formulated as
\begin{equation}
G_{\text{retr}}\!\!:\!
\begin{cases}
\mathbf{h}_i^{\text{layout}} = \mathrm{GT}_1(\mathbf{x}_i^{\text{phys.}} + \mathbf{x}_i^{\text{struct.}}), 
\mathbf{E}_{\text{layout}} = \mathrm{Pool}\!\left(\{\mathbf{h}_i^{\text{layout}}\}_i\right)\\
\mathbf{h}_j^{\text{logic}} = \mathrm{GT}_2(\mathbf{x}_j^{\text{struct.}}),
\mathbf{E}_{\text{logic}} = \mathrm{Pool}\!\left(\{\mathbf{h}_j^{\text{logic}}\}_j\right)
\end{cases}
\end{equation}
During retrieval, we compute the cosine similarity between the layout embedding query $\mathbf{E}_{\text{layout}}$ and each logic embedding $\mathbf{E}_{\text{logic}}^{(k)}$  from the embedding pool to rank the equivalent logic structures:
\begin{equation}
\text{Retrieve: }\text{Sim}_{\cos}(\mathbf{E}_{\text{layout}}, \mathbf{E}_{\text{logic}}^{(k)}) = 
\frac{\mathbf{E}_{\text{layout}} \cdot \mathbf{E}_{\text{logic}}^{(k)}}
{\|\mathbf{E}_{\text{layout}}\|_2 \, \|\mathbf{E}_{\text{logic}}^{(k)}\|_2}
\end{equation}
\textbf{Physical attribute generation: cell-level decoder.}
After retrieving the new logic structures, we generate their cell-level physical attributes using two decoder heads. For each gate, the decoder takes a combined feature $\mathbf{h}_j^{\text{dec}}$ as input. It consists of 1) the retrieved logic embedding $\mathbf{h}_j^{\text{logic}}$, 2) the layout MFFC constraints $\mathbf{x}_\text{con}$ (i.e., its IO net slack and slew), and 3) the layout MFFC embedding $\mathbf{E}_{\text{layout}}$.
This combined feature conditions the cell decoder to generate sizes and locations that remain consistent with the surrounding physical context, ensuring that the rewritten structure integrates smoothly into the existing layout.
% The decoder is implemented as two MLPs, for sizing classification and location regression, respectively. The sizing head performs masked classification: it outputs logits over all size options and then applies a legality mask based on the cell type so that the softmax distribution and final choice are restricted to library-valid sizes only. The location head performs normalized regression: it predicts $(u_x,u_y)\in[-1,1]^2$ for each gate, which are then linearly mapped into the original MFFC bounding box to obtain physical coordinates, ensuring all predicted locations stay within a feasible local region while allowing flexible placement adjustments. 

The decoder uses two MLP heads taking $\mathbf{h}_j^{\text{dec}}$ as input for sizing classification and location regression, respectively. The sizing head $\mathrm{MLP}_{\text{size}}$ performs masked classification. It predicts logits over all size options and applies a legality mask based on each cell type so that final predictions are restricted to library-valid sizes only. The location head $\mathrm{MLP}_{\text{loc}}$ performs normalized regression. It predicts the normalized coordinates $(\tilde{x}_j,\tilde{y}_j) \in [-1,1]^2$ for each gate, which are linearly mapped to the original MFFC bounding box to obtain physical coordinates, ensuring all placements remain within a feasible region while still allowing flexible adjustments. The generation process is formulated as follows:
\begin{equation}
G_{\text{gen}}\!:\!
\begin{cases}
\mathbf{h}_j^{\text{dec}} = \big[\mathbf{h}_j^{\text{logic}},\, \mathbf{x}_\text{con},\, \mathbf{E}_{\text{layout}}\big]
& (\text{Gate feat.}) \\[4pt]
p(s_j \mid \mathbf{h}_j^{\text{dec}}) = \mathrm{Softmax}\!\left(\mathrm{MLP}_{\text{size}}\!\left(\mathrm{mask}(\mathbf{h}_j^{\text{dec}})\right)\right)
& (\text{Size}) \\[4pt]
(\tilde{x}_j,\tilde{y}_j) = \mathrm{MLP}_{\text{loc}}\!\left(\mathbf{h}_j^{\text{dec}}\right),\;\; (\tilde{x}_j,\tilde{y}_j)\in[-1,1]^2
& (\text{Location})
\end{cases}
\end{equation}
This encoder–decoder graph RAG model bridges logic structure and physical attributes, enabling each layout MFFC to be rewritten into a new variant that can be accurately evaluated under layout timing. Moreover, the RAG model can be trained in a data-driven manner to jointly select alternative logic structures and generate corresponding cell sizes and locations, paving the way for our next step of RL-based timing optimization.

\begin{table*}[!t]
% \vspace{-.3in}
% \caption{Comparison of post-placement PPA results. E-Layout outperforms state-of-the-art layout restructuring tools}%applied in optimization.}
\caption{Comparison of post-placement optimization PPA results. E-Layout outperforms state-of-the-art layout restructuring tools, delivering larger timing improvements.}
\vspace{-.1in}
\resizebox{1\textwidth}{!}{

\begin{tabular}{c||ccc|ccc|ccc||ccc} \toprule
                                  & \multicolumn{3}{c|}{\textbf{Commercial Tool}}                                                                                                                                     & \multicolumn{3}{c|}{\textbf{w. OpenRoad-Restructure~\cite{ajayi2019openroad}}}                                                                                                        & \multicolumn{3}{c||}{\textbf{w. OpenPhySyn-Restructure~\cite{agiza2020openphysyn}}}                                                                                                      & \multicolumn{3}{c}{\cellcolor[HTML]{cde09a}\textbf{w. E-Layout (ours)}}                                                                                                             \\ \cline{2-13}
\multirow{-2}{*}{\textbf{Design}} & \begin{tabular}[c]{@{}c@{}}\textbf{TNS} \\      \small$(ns)$\end{tabular} & \begin{tabular}[c]{@{}c@{}}\textbf{WNS} \\      \small$(ns)$\end{tabular} & \begin{tabular}[c]{@{}c@{}}\textbf{Area} \\      (\small$mm^2$)\end{tabular} & \begin{tabular}[c]{@{}c@{}}\textbf{TNS} \\      \small$(ns)$\end{tabular} & \begin{tabular}[c]{@{}c@{}}\textbf{WNS} \\      \small$(ns)$\end{tabular} & \begin{tabular}[c]{@{}c@{}}\textbf{Area} \\      (\small$mm^2$)\end{tabular} & \begin{tabular}[c]{@{}c@{}}\textbf{TNS} \\      \small$(ns)$\end{tabular} & \begin{tabular}[c]{@{}c@{}}\textbf{WNS} \\      \small$(ns)$\end{tabular} & \begin{tabular}[c]{@{}c@{}}\textbf{Area} \\      (\small$mm^2$)\end{tabular} & \begin{tabular}[c]{@{}c@{}}\textbf{TNS} \\      \small$(ns)$\end{tabular} & \begin{tabular}[c]{@{}c@{}}\textbf{WNS} \\      \small$(ns)$\end{tabular} & \begin{tabular}[c]{@{}c@{}}\textbf{Area} \\      (\small$mm^2$)\end{tabular} \\ \midrule
s838\_1\_bench                    & -1.93                                                    & -0.108                                                   & 46                                                         & { -3.1 (60.62\%)}                    & { -0.184}                  & 41                                                         & { -1.891 (-2.02\%)}                  & { -0.102}                  & 45                                                         & { -1.99 (3.02\%)}                    & { -0.11 (1.82\%)}                    & 45                                                         \\
wb\_dma\_top                      & -47.76                                                   & -0.103                                                   & 426                                                        & { -67 (40.28\%)}                     & { -0.158}                   & 386                                                        & { -45.5 (-4.73\%)}                   & { -0.11}                     & 424                                                        & { -42.33 (-12.83\%)}                 & { -0.101 (-1.98\%)}                    & 372                                                        \\
spi\_top                          & -31.92                                                   & -0.192                                                   & 259                                                        & { -31.2 (-2.26\%)}                   & { -0.154}                 & 240                                                        & { -31.3 (-1.94\%)}                   & { -0.192}                      & 267                                                        & { -30.861 (-3.43\%)}                 & { -0.191 (-0.52\%)}                  & 268                                                        \\ \cline{5-10}  
usbf\_top                         & -13.1                                                    & -0.126                                                   & 1070                                                       & \multicolumn{6}{c||}{N/A$\star$}                                                                                                                                                                                                                                                                                                                                             & { -11.268 (-16.26\%)}                & { -0.108 (-16.67\%)}                 & 1041                                                       \\ \cline{5-10}  
NV\_csb\_master                & -77.1                                                    & -0.1                                                     & 1253                                                       & { -72 (-6.61\%)}                     & { -0.117 }                     & 1235                                                       & { -73 (-5.32\%)}                     & { -0.093 }                     & 1253                                                       & { -70.32 (-9.6\%)}                  & { -0.105 (4.76\%)}                   & 1282                                                       \\ \cline{5-10}  
NV\_bdma                       & -58.1                                                    & -0.111                                                   & 539                                                        & \multicolumn{6}{c||}{N/A$\star$}                                                                                                                                                                                                                                                                                                                                           & { -59 (1.53\%)}                      & { -0.112 (0.89\%)}                   & 547                                                        \\  \cline{5-10}  
ac97\_top                         & -92.5                                                    & -0.085                                                   & 1183                                                       & { -107 (15.68\%)}                    & { -0.107}                  & 1181                                                       & { -94 (1.62\%)}                      & { -0.092}                   & 1177                                                       & { -83.1 (-11.31\%)}                  & { -0.085 (0\%)}                      & 1193                                                       \\ \cline{5-10}  
arm\_core                         & -558                                                     & -0.594                                                   & 2196                                                       & \multicolumn{6}{c||}{}                                                                                                                                                                                                                                                                                                                                                & { -550 (-1.45\%)}                    & { -0.564 (-5.32\%)}                  & 2271                                                       \\
Rocket                            & -555                                                     & -0.405                                                   & 1378                                                       & \multicolumn{6}{c||}{\multirow{-2}{*}{N/A$\star$}}                                                                                                                                                                                                                                                                              & { -537.6 (-3.24\%)}                  & { -0.391 (-3.58\%)}                  & 1360                                                       \\ \midrule
\textbf{Avg.}                     &                                                          &                                                          &                                                            & { \color[HTML]{FF0000}\textbf{21.54\%}}                  & { \color[HTML]{FF0000}\textbf{29.37\%}}                  & { \color[HTML]{00B050}\textbf{-5.82\%}}                    & { \color[HTML]{00B050}\textbf{-1.37\%}}                  & { \color[HTML]{FF0000}\textbf{0.50\%}}                   & { \color[HTML]{00B050}\textbf{0\%}}                        & { \color[HTML]{00B050}\textbf{-6.07\%}}                  & { \color[HTML]{00B050}\textbf{-2.29\%}}                  & { \color[HTML]{00B050}\textbf{-1.07\%}}          \\ \bottomrule          
\end{tabular}

}
\begin{tablenotes}\footnotesize
\item $^\star$ N/A indicates cases where the corresponding open-source flow cannot support multi-output gates, so the design cannot be processed under that baseline.
\end{tablenotes}

\label{tbl:res}
\vspace{-.1in}
 
\end{table*}

\subsection{Tuning Graph RAG with RL for Global Timing Optimization}\label{sec:rl}

The key challenge is that layout-stage restructuring should be optimized for \emph{design-level timing}, yet no direct supervision exists for this objective. There is no practical ground-truth label for the best logic structure, gate sizes, and gate locations of each local subcircuit once full-chip timing interaction is taken into account. Existing tool solutions are also largely local and heuristic, and therefore do not provide suitable supervision for globally optimal restructuring.

We therefore further optimize the graph RAG model with reinforcement learning using full-chip TNS as reward feedback. This allows E-Layout to refine logic structure selection and physical realization across multiple MFFCs according to their actual design-level timing impact. In this way, RL provides the link between local restructuring decisions and the global post-placement timing objective.

\textbf{GRPO sampling and reward design.}
We adopt Group Relative Policy Optimization (GRPO)~\cite{shao2024deepseekmath, guo2025deepseek}, a lightweight policy-gradient reinforcement learning method. Its key idea is to sample multiple candidate actions for the same state, evaluate them with a task-level reward, and update the ML model according to their \emph{relative} performance within the group. In our setting, the graph RAG model refers to the policy, and layout timing improvement defines the reward. For each layout MFFC, denoted as state $s$, we sample a group of $G$ candidate restructurings. Each candidate is a joint action $a^{(g)}$ drawn from three model distributions: the structure-retrieval distribution $D_\text{struct.}$ from $G_{\text{retr}}$, and the sizing and location distributions $D_\text{size}$ and $D_\text{loc.}$ from $G_{\text{gen}}$.

The sampled candidates replace the original layout MFFC to form $G$ new layouts, which are evaluated by the fast GPU-accelerated incremental STA engine INSTA~\cite{luinsta}. To capture interactions among different MFFCs under full-chip timing, we sample and update candidates over a mini-batch of MFFCs. Based on their timing outcomes, we then compute group-wise relative advantages as:
\begin{equation}
A^{(g)} = \frac{\Delta \mathrm{TNS}^{(g)} - \mu_{\mathrm{TNS}}}{\sigma_{\mathrm{TNS}}},
\end{equation}
where $\Delta \mathrm{TNS}$ denotes the improvement in total negative slack. Each objective is standardized independently to maintain scale invariance. GRPO uses the mean advantage within each candidate group as an implicit baseline, eliminating the need for an explicit value-function critic or external ranking model and thus reducing computational overhead. 

\textbf{GRPO training iteration.}
During GRPO training, we maintain a stochastic graph RAG model $\pi_{\theta}$ and a frozen reference graph RAG model $\pi_{\mathrm{ref}}$. In each iteration, we sample a mini-batch of MFFCs and use $\pi_{\theta}$ to generate a group of candidate restructuring solutions for each MFFC, as described in the previous paragraph. These candidates are substituted back into the layout and evaluated by INSTA to obtain timing slack improvement $\Delta \mathrm{TNS}$, which is then normalized to compute the group-relative advantage $A^{(g)}$. Finally, we update $\pi_{\theta}$ by maximizing a clipped surrogate objective:
\begin{equation}
\begin{aligned}
J_{\text{GRPO}}(\theta) =
\mathbb{E}\!\Big[
& \min\!\big(r^{(g)} A^{(g)},\, 
\mathrm{clip}(r^{(g)}, 1-\epsilon, 1+\epsilon) A^{(g)}\big) \\
& - \beta D_{\mathrm{KL}}\!\big(\pi_{\theta}\,\|\,\pi_{\mathrm{ref}}\big)
\Big],
\end{aligned}
\end{equation}
where $r^{(g)}$ is the importance sampling ratio that measures how much the new model $\pi_{\theta}$ differs from the reference old model $\pi_{\mathrm{ref}}$ for the same sampled action $a^{(g)}$, defined as $r^{(g)} = \frac{\pi_\theta(a^{(g)} \mid s)}{\pi_{\mathrm{ref}}(a^{(g)} \mid s)}$, and  $D_{\mathrm{KL}}$ is the KL divergence between $\pi_{\theta}$ and $\pi_{\mathrm{ref}}$, ensuring that updates do not push the model too far away from the reference behavior.

% \begin{equation}
% r^{(g)} = \frac{\pi_\theta(a^{(g)} \mid s)}{\pi_{\mathrm{ref}}(a^{(g)} \mid s)},
% \qquad
% D_{\mathrm{KL}}(\pi_\theta \,\|\, \pi_{\mathrm{ref}})
% = \mathbb{E}_{a \sim \pi_\theta}
% \left[\log \frac{\pi_\theta(a \mid s)}{\pi_{\mathrm{ref}}(a \mid s)}\right].
% \end{equation}

We partition all MFFCs into $m$ mini-batches, and apply GRPO on each batch for $n$ iterations\footnote{In practice, we set $m=10\sim20$ and $n=5\sim10$ to balance performance and runtime.}, keeping the best restructuring solution before moving to the next batch. After all MFFCs are processed, the resulting restructured layout is passed back to the commercial flow for incremental legalization and further buffering and sizing.

% \vspace{-.1in}
\section{Experimental Results} \label{sec:expr}
% \subsection{Experimental Setup}

\begin{figure*}[!t]
  \centering
  \vspace{-.4in}
  \includegraphics[width=1\linewidth]{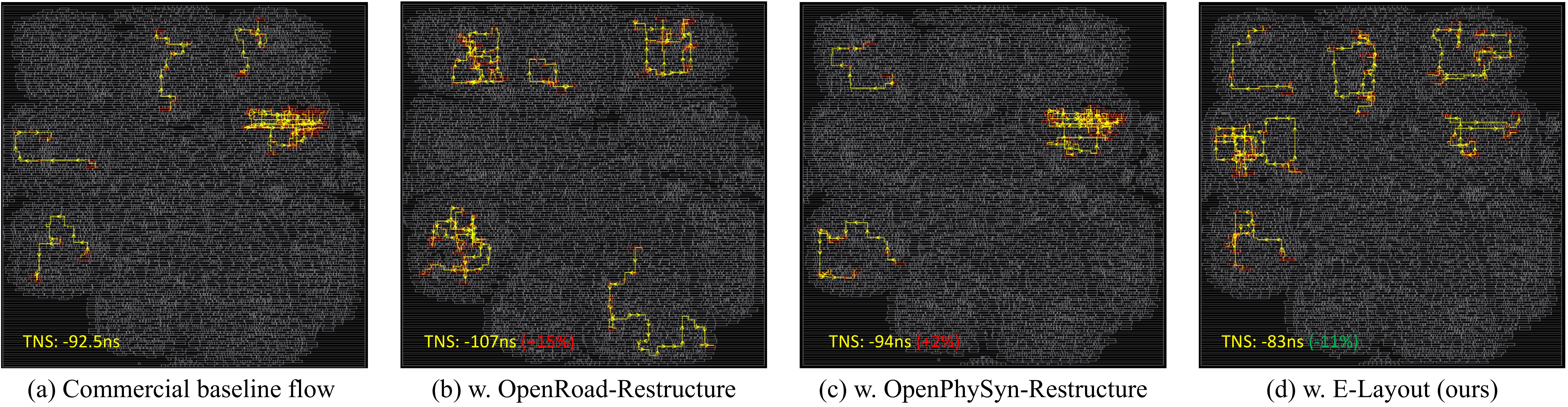}
  \vspace{-.25in}
  \caption{Layout visualization of different methods. Top-20 critical paths are highlighted. Compared with the commercial baseline (a) and open-source restructuring flows (b)–(c), our E-Layout (d) delivers better timing results with fewer long detours.\looseness=-1} 
  \label{fig:vis2}
  \vspace{-.1in}
\end{figure*}

\subsection{Experimental Setup}

\textbf{Dataset and evaluation setup.}
We evaluate E-Layout on 9 diverse open-source designs from OpenCores~\cite{URL:opencore}, ITC’99~\cite{corno2000rt}, Chipyard~\cite{amid2020chipyard}, and NVDLA~\cite{nvdla}. All experiments are conducted using the ASAP 7\,nm technology library within a leading commercial physical design flow. After E-Layout restructuring, the restructured layout is passed back to the commercial tool for incremental legalization and subsequent structure-invariant optimization, so the reported PPA reflects the effect of E-Layout after downstream placement optimization. We evaluate post-optimization TNS, WNS, and area. To ensure correctness, we employ formal verification at both the MFFC and full-chip levels: for MFFCs, we implement a combinational equivalence checker for standard-cell graphs using an SMT solver~\cite{gario2015pysmt}, and for full designs, we use Synopsys Formality for equivalence checking. During training and candidate evaluation, INSTA~\cite{luinsta} is used to provide fast incremental layout TNS feedback.

\textbf{E-Layout implementation.}
E-Layout is implemented using \texttt{egg-log}~\cite{zhang2023better} for e-graph equality saturation. For the graph RAG model, we use SGFormer~\cite{wu2023sgformer} as the graph-transformer backbone for the retrieval encoder model, producing 768-dimensional embeddings, followed by MLP heads for sizing classification and location regression. The graph RAG model is further tuned with reinforcement learning using full-chip timing feedback. We adopt MFFCs as the basic restructuring granularity, and only optimize MFFCs with more than 5 equivalent logic candidates, thereby focusing optimization on cones with meaningful structural alternatives.

\textbf{Baseline setup.}
We compare E-Layout against representative open-source layout restructuring flows, including OpenROAD~\cite{ajayi2019openroad} and OpenPhySyn~\cite{agiza2020openphysyn}. To ensure fair comparison, all baselines start from the same post-placement design state and are followed by the same downstream commercial legalization and optimization flow as E-Layout.
For OpenROAD~\cite{ajayi2019openroad}, it identifies timing-critical regions and invokes ABC~\cite{brayton2010abc} for logic remapping, without physical-awareness. For OpenPhySyn, we enable its default restructuring-related transforms, including pin swapping and gate cloning.

\subsection{Optimization Results}

\textbf{E-Layout restructuring results.}
\Cref{tbl:res} summarizes the post-placement PPA results across nine designs. E-Layout achieves on average 6.07\% TNS improvement, 2.29\% WNS improvement, and 1.07\% area reduction over the commercial post-placement baseline. These gains are notable because the starting point has already undergone substantial commercial timing optimization, leaving limited room for further improvement.

\textbf{Comparison with open-source layout restructuring tools.}
Compared with existing open-source restructuring flows, E-Layout delivers consistently stronger timing gains. OpenROAD-Restructure first identifies timing-critical regions after placement, but then delegates restructuring to ABC, whose logic remapping is performed without true layout awareness. As a result, the rewritten structures often break physical coherence with the surrounding layout and exhibit weak, or even negative, post-placement correlation, leading to average TNS degradation of up to 21.54\%. OpenPhySyn-Restructure is more physically aware, but its optimization is limited to local cell-level edits such as pin swapping, gate cloning, buffering, and resizing. Because it does not explore meaningfully different logic structures, its improvement space remains narrow, resulting in 1.37\% average TNS degradation and almost no WNS or area benefit. In contrast, E-Layout jointly optimizes equivalence-preserving logic restructuring and physical realization under full-chip timing feedback, consistently improving both TNS and WNS while maintaining competitive area.

Some baseline entries are marked as unavailable because the corresponding open-source flows cannot support specific design constructs in our setup, such as multi-output gates. Nevertheless, the overall trend is clear: E-Layout provides the strongest and most consistent post-placement optimization among all compared methods.\looseness=-1

\begin{table}[!t]

\caption{Statistics of E-Layout Restructuring.}%applied in optimization.}
\vspace{-.15in}
\resizebox{0.47\textwidth}{!}{

\begin{tabular}{l|c||l|c} \toprule
\multicolumn{2}{c||}{\textbf{Design-level}} & \multicolumn{2}{c}{\textbf{MFFC-level}} \\ \midrule
\#. Cells          & \{423, 4320, 8257\}  & \#. Restr. MFFCs      & \{17, 111, 200\}   \\
\#. Restr. cells      & \{111, 263, 430\}    & - \#. Expansion    & \{10, 80, 152\}    \\
- Percentage       & \{4\%, 9\%, 26\%\}   & - \#. Merge        & \{0, 3, 6\}        \\
\#. Cells in MFFC  & \{1, 3, 27\}         & - \#. Size only    & \{5, 28, 53\}   \\ \bottomrule  
\end{tabular}

}
\begin{tablenotes}\footnotesize
\item These statistics are represented in format \{min, avg, max\}.
\end{tablenotes} 

\label{tbl:statis}
\vspace{-.2in}
 
\end{table}

\subsection{Restructure Statistics}
\Cref{tbl:statis} summarizes the restructuring behavior of E-Layout. We only optimize MFFCs with more than 5 equivalent structures, among which about 9\% are eventually rewritten. Each rewritten MFFC contains 3 cells on average. Although these cones are small, they are composed of complex standard cells rather than simple primitive gates (e.g., AIG), and can therefore admit rich combinational flexibility. In practice, even an MFFC with only a few gates can expand into many tens or even hundreds of equivalent alternatives after equality saturation. This makes such cones meaningful restructuring units in the post-placement setting: they are local enough to preserve physical coherence, yet still large enough to expose substantial structural diversity. Among the rewritten cones, roughly 70\% are expanded from complex cells into compositions of simpler gates, 3\% are merged into more complex cells, and the remaining 27\% preserve the logic structure while only gate sizes are changed.

\subsection{Case Study and Visualization}

\begin{figure}[!t]
  \centering
  % \vspace{-.3in}
  \includegraphics[width=1\linewidth]{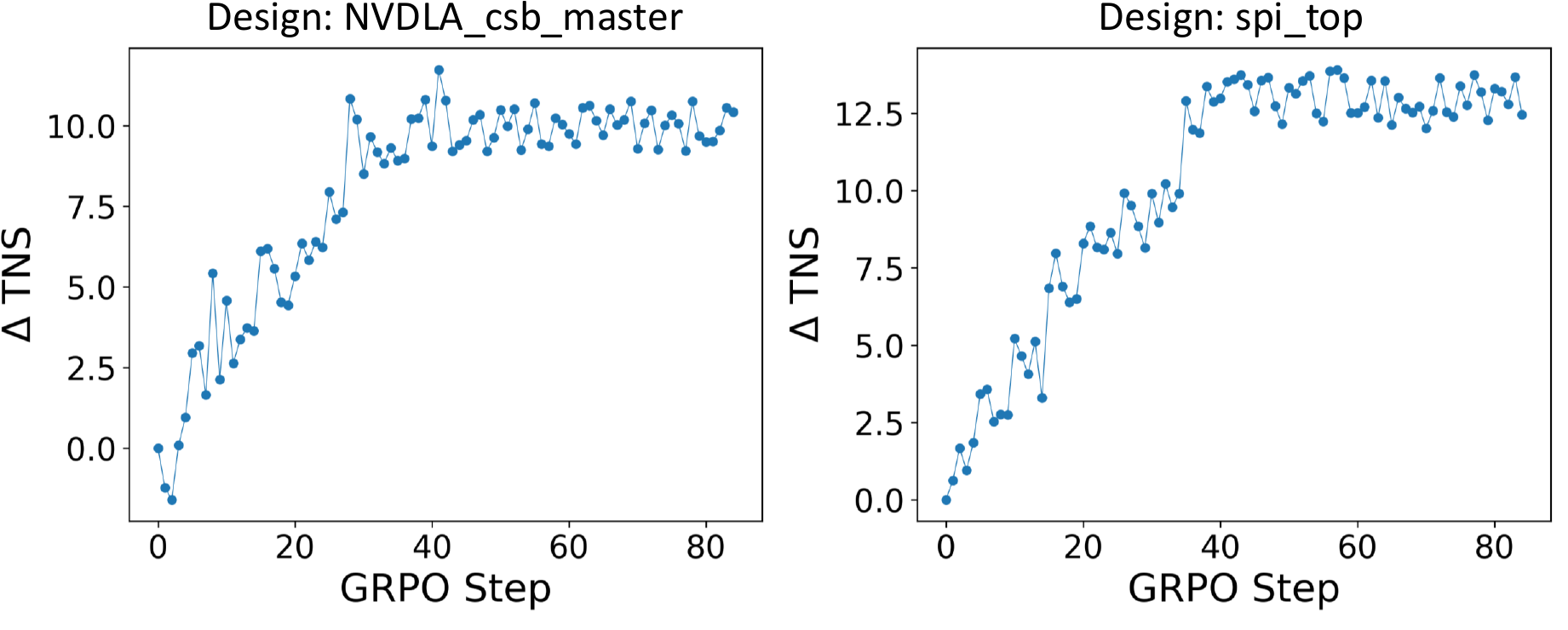}
  \vspace{-.15in}
  \caption{GRPO training curves demonstrating consistent TNS improvements over GRPO steps.\looseness=-1} 
  \label{fig:vis3}
  \vspace{-.2in}
\end{figure}

\textbf{RL tuning trajectory.}
\Cref{fig:vis3} shows the GRPO training trajectories in terms of $\Delta\text{TNS}$ evaluated by INSTA. Every 5 GRPO steps, E-Layout optimizes a batch of MFFCs, selects the best restructuring solution, and then moves to the next batch. After one full pass over all candidate MFFCs, the second epoch converges quickly, indicating that the learned policy stabilizes and that the remaining timing improvement space becomes progressively smaller.

\textbf{Design-level visualization.}
\Cref{fig:vis2} visualizes the top-20 critical paths for all methods after post-placement optimization. The commercial baseline and open-source restructuring flows still exhibit long, circuitous detours. Although OpenROAD performs aggressive remapping, its ABC-based optimization is not layout-aware and can even degrade timing. In contrast, E-Layout produces more direct critical paths and alleviates detours in dense regions, suggesting that the selected restructurings are better aligned with the surrounding physical context.

\begin{figure}[!t]
  \centering
  % \vspace{-.3in}
  \includegraphics[width=1\linewidth]{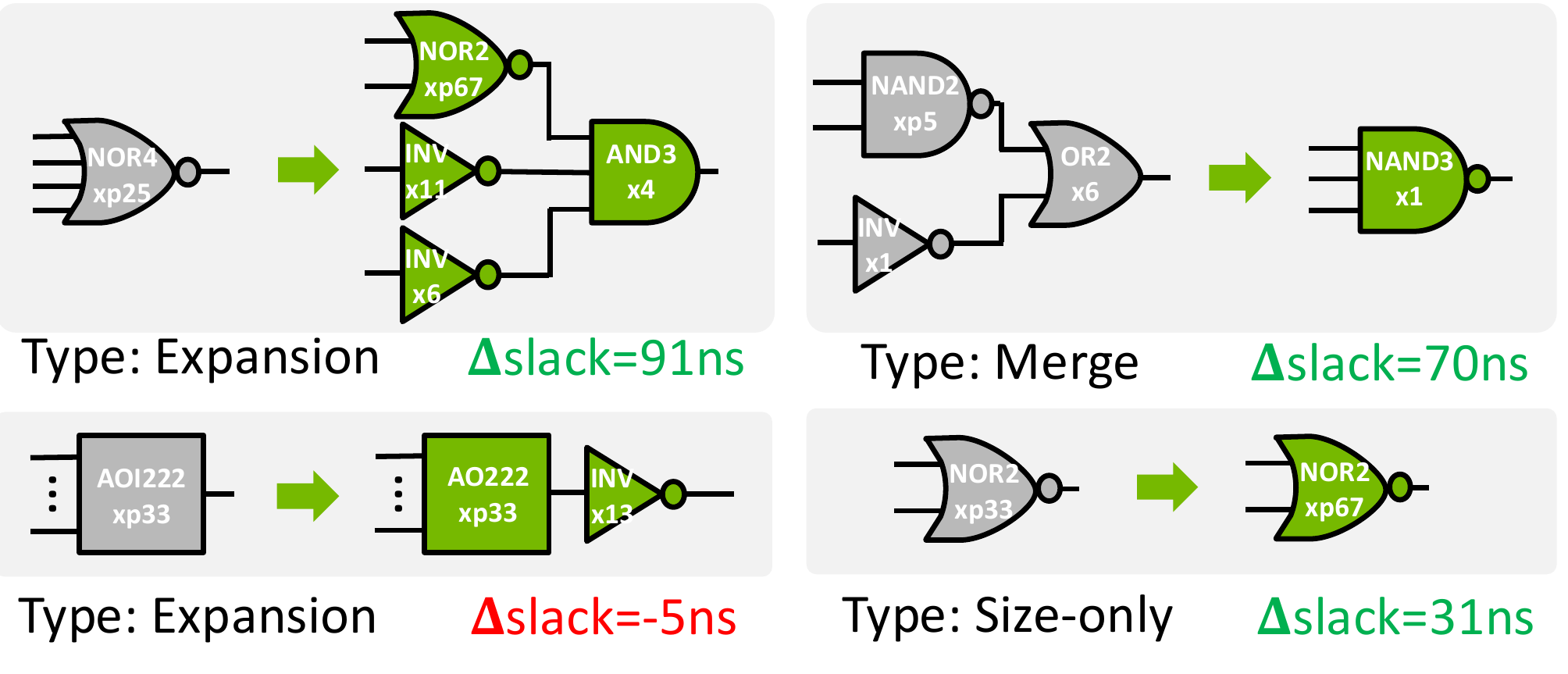}
  \vspace{-.1in}
  \caption{Case study of different MFFC restructuring types: expansion, merging, and sizing-only} 
  \label{fig:vis}
  % \vspace{-.1in}
\end{figure}

\textbf{MFFC-level restructuring case studies.}
\Cref{fig:vis} illustrates three representative cone-level decisions in E-Layout: expansion of a complex cell into simpler gates, merging of multiple gates into a more complex cell, and size-only tuning. These examples show that E-Layout does not favor a single restructuring pattern; instead, it adaptively chooses among structural expansion, merging, and physical-only adjustment according to local structure, physical context, and design-level timing feedback.

\subsection{Ablation Study}
\begin{figure}[!t]
  \centering
  \includegraphics[width=0.75\linewidth]{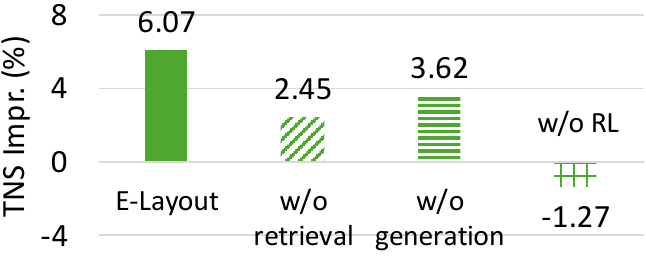}
  \vspace{-.1in}
  \caption{Ablation study of E-Layout ML framework on logic structure retrieval, physical attribute generation, and RL tuning for the graph RAG model.}
  \label{fig:abl}
  % \vspace{-.1in}    
\end{figure}

We conduct ablation studies to evaluate the contribution of each component in the E-Layout framework, with results summarized in~\Cref{fig:abl}. Removing the retrieval module and reducing the method to sizing-only optimization lowers the average TNS gain by about 3.5\%, showing that selecting among equality-saturated logic alternatives is essential rather than merely resizing the original structure. Disabling the generation by using median sizes and random locations causes a further 2.4\% drop, indicating the importance of jointly optimizing physical realization rather than treating it as a separate downstream step. Finally, removing RL fine-tuning and relying only on supervised tuning with local tool restructuring labels leads to the largest degradation, even worsening timing, which underscores the importance of full-chip timing feedback for aligning local restructuring decisions with the global objective.

\subsection{Runtime and Scalability Analysis}
\Cref{tbl:runtime} summarizes the runtime decomposition and scalability of E-Layout. The total runtime consists of two parts: local preprocessing and iterative optimization. The preprocessing stage includes netlist-to-graph conversion, MFFC extraction, and e-graph saturation. These steps are local to each MFFC, scale with design size and cone complexity, and are naturally parallelizable across CPU cores.

The iterative optimization stage includes graph RAG inference, RL tuning, and INSTA timing evaluation. Graph RAG inference is lightweight, completing in less than one second per batch on a single GPU, whereas RL tuning scales with the number of iterations and group size. In practice, incremental INSTA timing evaluation dominates the optimization loop. Overall, E-Layout follows a \emph{local generation, global evaluation} workflow: local candidate generation scales well under parallel execution, while design-level timing evaluation is the main bottleneck, resulting in total runtime from minutes to hours depending on design size and optimization iterations.\looseness=-1

\vspace{-.05in}

\section{Discussion}

\textbf{Graph RAG as a learned extraction mechanism for e-graphs.}
Traditional e-graph optimization typically relies on rule-driven extraction with simple cost functions, such as area or delay, which are insufficient to capture the coupled logic--physical dependencies of post-layout timing. In E-Layout, the graph RAG model serves as a learned extraction mechanism: instead of selecting structures with a fixed linear cost, it retrieves promising logic alternatives and generates their physical realization conditioned on both logic and layout context. This allows e-graph extraction to be guided by richer logic--physical correlations than conventional heuristic cost functions.\looseness=-1

\textbf{Optimization granularity.}
The choice of MFFCs as the restructuring granularity reflects a practical tradeoff between optimization scope and physical controllability. If the region is too small, the available structural flexibility quickly collapses into sizing-like local adjustment; if it is too large, restructuring becomes increasingly entangled with neighboring logic and surrounding layout, making physically coherent implementation much harder. MFFCs occupy a useful middle ground: they are large enough to expose nontrivial logic alternatives, yet local enough to support incremental legalization and preserve the coherence of the surrounding placement. In principle, the same framework could be extended to finer or coarser regions, but doing so would require explicitly modeling stronger interaction across restructurings, especially for overlapping or mutually dependent optimization regions.

\textbf{Toward unified layout optimization.}
At present, E-Layout focuses on ML-driven restructuring, while other layout-stage optimizations, such as buffering and sizing, are still handled by commercial tools. As a result, the final PPA can only be observed after the complete downstream flow, which makes closed-loop optimization relatively expensive. A natural next step is to move toward a unified framework in which restructuring, buffering, and sizing are optimized jointly under a shared timing objective. This could further reduce mismatch across stages and enable more direct optimization toward final layout quality.

\begin{table}[!t]

\caption{Runtime and scalability analysis of E-Layout.}%applied in optimization.}
\vspace{-.05in}
\resizebox{0.5\textwidth}{!}{

\begin{tabular}{l|lllc} \toprule
\textbf{Stage}      & \textbf{Complexity / Scalability}         & \textbf{Runtime}             & \textbf{Parallel} &  \\ \midrule \midrule
Netlist-graph       & Linear in number of gates                 & sec-min           &                         &  \\
MFFC extract     & Linear in MFFC gate number                & sec-min           & \Checkmark                       &  \\
E-graph rewrite  & Scales with cone size/ equivalent space & sec-min           & \Checkmark                       &  \\
RAG inference & Batched GPU inference & \textless 1s/batch & \Checkmark                       &  \\
RL tuning           & Scales with iterations and group size  & sec-min           &                         &  \\
INSTA eval.    & Scales with  design size                  & sec-min           & \Checkmark                       &  \\ \midrule
Overall             & Scales with  design size and   iterations & min-hours            &                         &  \\ \bottomrule
\end{tabular}

}

\label{tbl:runtime}
% \vspace{-.1in}
 
\end{table}

% E-Layout complements current AI-assisted layout optimizations that focus on structure-invariant techniques such as generative buffering and sizing. By introducing equivalence-preserving structural rewrites at the layout stage, it explores a broader design space that conventional buffering and sizing inherently cannot reach. Moreover, E-Layout can operate as an upstream enhancement: the restructured netlist can still benefit from downstream ML-based buffering and sizing, enabling a more holistic AI-augmented physical design pipeline. 
% A current limitation is that our RL training loop uses timing feedback only before these downstream optimizations; the final post-optimization timing is not yet incorporated into the reward. Integrating restructuring, buffering, and sizing under a unified end-to-end timing objective remains an important direction for future work.

\section{Conclusion}\label{sec:concl}

We presented E-Layout, an ML-driven framework for post-placement layout restructuring. E-Layout preserves functional equivalence while jointly optimizing logic structure and physical realization under accurate layout timing. Its central contribution is to formulate layout-stage restructuring as a coupled logic--layout co-optimization problem spanning equivalence-preserving structure exploration, logic selection with physical realization, and design-level timing-driven refinement. By integrating e-graph equality saturation, graph RAG, and reinforcement learning with full-chip feedback, E-Layout enables structure-changing optimization directly at layout stage beyond conventional structure-invariant post-placement optimization. Evaluated on a 7\,nm commercial physical design flow, E-Layout achieves on average 6\% TNS improvement, 2\% WNS improvement, and 1\% area reduction, while consistently outperforming existing open-source layout restructuring tools.

\section{ACKNOWLEDGEMENT}
This work is supported by Hong Kong Research Grants Council (RGC) CRF-YCRG C6003-24Y, 16216825, and T46-415/25-R. It is also supported by RGC JLFS-YSF/E-605/26. It was partially conducted by ACCESS – AI Chip Center for Emerging Smart Systems, supported by the InnoHK initiative of the Innovation and Technology Commission of the Hong Kong Special Administrative Region Government.

\clearpage
\bibliographystyle{ACM-Reference-Format}
\bibliography{ref}

\end{document}